# *Information and Stock Prices:*
# *A Simple Introduction*

## Kenton K. Yee[*]

## Columbia University

[*] Graduate School of Business, Columbia University.

Mail: Columbia University, 615 Uris Hall, 3022 Broadway, New York, NY 10027
URL: http://www.columbia.edu/~kky2001/

# Information and Stock Prices:
# A Simple Introduction

This article summarizes recent research in financial economics about why information, such as earnings announcements, moves stock prices. The article does not presume any prior exposure to finance beyond what you might read in newspapers.

"Quality information is the lifeblood of strong, vibrant markets. Without it, investor confidence erodes. Liquidity dries up. Fair and efficient markets simply cease to exist. As the *quantity* of information increases exponentially through the Internet and other technologies, the *quality* of that information must be our signal priority."

[SEC Chairman Arthur Levitt (1999)]

*The No-Arbitrage Principle*

As every casual follower of financial news knows, stock prices rise and fall in response to earnings and revenues. Almost every AP newswire or CNBC report on quarterly financial announcements focuses on two questions: whether a firm is meeting or beating earnings per share (EPS) expectations, and whether management is raising or lowering projections for future revenues. There is little discussion of cash flows or changes in the value of important balance sheet items like real property, which may be worth billions of dollars (Yee [1]).

On its face, this observation appears to be at odds with the no-arbitrage principle—the basis of the traditional approach to asset pricing in finance—which states that cash flows determine equilibrium prices. According to this principle, shareholders should pay only for expected real cash flows, not bookkeeping constructs like earnings.

As any follower of financial news knows, earnings is loaded with bookkeeping accruals—depreciation, amortization, deferred taxes and valuation allowances— and so can deviate markedly from past, present or future cash flows even for stable, mature companies. Moreover, since investors rely on managers to estimate and report this number, it is subject to misreporting and manipulation.

*Why, then, do earnings announcements move markets, while cash flow news is all but ignored?*

### *The Complementarity Principle*

This state of affairs suggests that there may be a glaring discrepancy between traditional finance theory and how stock prices are set in real markets (Yee [1]). Intrigued by this problem, I set out to answer the following questions: Under what conditions is EPS a more potent mover of prices than information about cash flows? When do cash flow and balance sheet information gain in potency? And how does the quality of accounting affect the relative potency of EPS, cash flow, and balance sheet information?

The central tenet of this idea is the complementarity principle—the idea that future cash flows are predicted by a weighted average of the three main components of financial reports: balance sheet information, current cash flows, and forward EPS. At the heart of this idea is that these three financial components do not simply measure historical performance – they *forecast* future performance (Yee [2], [3]). That is, balance sheet information, current cash flows, and forward EPS are complementary forecasting attributes that predict the future cash flows priced by the no arbitrage principle. The complementarity principle links all three forecasting attributes to a company's stock price. In the model, the weight coefficients of the three attributes add up to one. The three attributes are complementary in the sense that if, for instance, EPS increases in significance, cash flow and balance sheet information become commensurately less important.

This framework helps to explain why investors and analysts sometimes rely more on cash flows and less on EPS and *vice versa*. The price of a mature, "cash cow" firm is essentially determined by capitalized cash flows; earnings and balance sheet information provide little additional information that would improve the valuation estimate of such a firm. A young growth firm, on the other hand, has cash flows that are negative and erratic, so it makes no sense to capitalize them; thus earnings and balance sheet information are the key valuation inputs. For growth firms, the relative importance of earnings versus balance sheet information depends on the expected earnings growth rate, earnings quality, the probability of bankruptcy, on and off balance sheet financial and operating leverage, and the market discount rate.

The complementarity principle suggests that the three attributes' weights automatically adjust to bookkeeping biases. A firm with aggressive revenue deferrals, for example, will have an inaccurately depressed value estimate based on capitalized earnings alone. In such a situation, balance sheet and cash flow information step in to "repair" any potential biases caused by capitalized earnings. I was surprised by how central a role bookkeeping accruals play, not only in valuation, but also in enabling analysts to forecast earnings more accurately.

The complementarity principle implies that the income statement, balance sheet, and statement of cash flows together provide the key to unlocking the mystery of stock prices and the risk-return relationship. In this sense, the Wall Street analysts who spend their time pouring over deferred tax footnotes, adjusting valuation allowances and fretting over arcane mergers and acquisitions accounting rules were on the right track all along.

*Concluding Remarks*

I am now investigating the possibility that the complementarity principle may hold the key to unlocking yet another important question: why do accounting ratios, like the market-to-book ratio, affect the risk-return relation in capital markets? My analysis shows that a poor-quality earnings estimate magnifies fundamental risk but cannot affect the market risk premium in the absence of fundamental risk. Thus investors and regulators want the highest quality accounting from the riskiest firms, such as those in speculative high-tech industries. Ironically, sometimes it is precisely those firms that have the worst earnings quality because of their off balance sheet activities, aggressive approach to employee stock option expenses, and strategic purchase price allocations in mergers and acquisitions accounting.

Building on these initial findings about the risk-return relation, I extended this model to distinguish between *content* quality and *precision* quality. Content quality refers to the amount of information earnings estimates contain about future cash flows, while precision quality refers to how precisely the estimates reflect that information. I showed that the complementarity principle prevails even when a firm has degraded both the content and precision quality of its financial statements. In such situations, the weight coefficients in the complementarity framework incorporate a structural factor related to earnings quality and accounting bias.

I went on to observe that the complementarity principle is remarkably consistent with everyday valuation practice in the court system. In a study of how Delaware Courts assess share value in shareholder lawsuits, I found that the courts employ a valuation formula that is essentially a special case of the complementarity principle: the judges collect earnings, balance sheet, and cash flow information, and then use a weighted average of market price and the collected accounting numbers to determine the value of minority shares (Yee [2], [4]). By rejecting market price as the final arbiter of share value, the judges are able to impose a minority discount, which provides incentives for controlling shareholders to make sound investment choices. Surprisingly, this method was developed by the state courts in the first half of the 20$^{th}$ century without any guidance from 21$^{st}$ century concepts like the complementarity principle.

## References

[1] Yee, K. "Earnings Quality and the Equity Risk Premium." *Contemporary Accounting Research*, forthcoming. PDF file available at
http://papers.ssrn.com/sol3/papers.cfm?abstract_id=846546

[2] Yee, K. "Perspectives: Combining Value Estimates to Increase Accuracy." *Financial Analysts Journal* 60, no. 4 (2004): 23–28.

[3] Yee, K. "Aggregation, Dividend Irrelevancy, and Earnings-Value Relations." *Contemporary Accounting Research* 22.2 (2005): 453-480. PDF file available at
http://papers.ssrn.com/sol3/papers.cfm?abstract_id=667781

[4] Yee, K. "Control Premiums, Minority Discounts, and Optimal Judicial Valuation." *Journal of Law and Economics* 48.2 (2005): 517-548.